\documentclass[
preprint,
 amsmath,amssymb,
 aps,
prb,
floatfix,
]{revtex4-2}

\usepackage{graphicx}
\usepackage{dcolumn}
\usepackage{booktabs, tabularx, ragged2e}
\usepackage{bm}
\usepackage[mathlines]{lineno}
\usepackage{commath}
\usepackage{physics}
\usepackage{color}
\usepackage{mathtools}

\begin{document}

\preprint{APS/123-QED}

\title{Emergence of Rashba-spin-valley state in two-dimensional strained bismuth oxychalcogenides Bi$_{2}$O$_{2}$Se}

\author{Muhammad Darwis Umar}
\affiliation{Departement of Physics, Faculty of Mathematics and Natural Sciences, Universitas Gadjah Mada, Sekip Utara BLS 21 Yogyakarta 55186 Indonesia.}

\author{Lalu Dalilul Falihin}
\affiliation{Departement of Physics, Faculty of Mathematics and Natural Sciences, Universitas Gadjah Mada, Sekip Utara BLS 21 Yogyakarta 55186 Indonesia.}

\author{Arief Lukmantoro}
\affiliation{Departement of Physics, Faculty of Mathematics and Natural Sciences, Universitas Gadjah Mada, Sekip Utara BLS 21 Yogyakarta 55186 Indonesia.}

\author{Harsojo}
\affiliation{Departement of Physics, Faculty of Mathematics and Natural Sciences, Universitas Gadjah Mada, Sekip Utara BLS 21 Yogyakarta 55186 Indonesia.}

\author{Moh. Adhib Ulil Absor}
\affiliation{Departement of Physics, Faculty of Mathematics and Natural Sciences, Universitas Gadjah Mada, Sekip Utara BLS 21 Yogyakarta 55186 Indonesia.}
\email{adib@ugm.ac.id}


\date{\today}

\begin{abstract}
The experimental evidence of the ultra-high electron mobility and strong spin-orbit coupling in the two-dimensional (2D) layered bismuth-based oxyselenide, Bi$_{2}$O$_{2}$Se, makes it a potential material for spintronic devices. However, its spin-related properties have not been extensively studied due to the centrosymmetric nature of its crystal structure. By using first-principles density-functional theory calculation, this study reports the emergence of Rashba-spin-valley states in Bi$_{2}$O$_{2}$Se monolayer (ML). Breaking the crystal inversion symmetry of Bi$_{2}$O$_{2}$Se ML using an external electric field enables the Rashba-spin-valley formation, causing the appearance of the Rashba-type splitting around the $\Gamma$ valley and spin-valley coupling at the $D$ valleys located near the middle of $\Gamma-M$ line. In addition to the typical Rashba-type spin textures around the $\Gamma$ valley, the study also observed in-plane unidirectional spin textures around the $D$ valleys, which is a rare phenomenon in 2D materials. The observed Rashba-spin-valley states are driven by the lowering point group symmetry of the crystal from $D_{4h}$ to $C_{4v}$ enforced by the electric field, as clarified through $\vec{k}\cdot\vec{p}$ model derived from symmetry analysis. More importantly, tuning the Rashba and spin-valley states by using biaxial strain offers a promising route to regulate the spin textures and spin splitting preventing the electron from back-scattering in spin transport. Finally, we proposed a more realistic system, namely, Bi$_{2}$O$_{2}$Se ML/SrTiO$_{3}$ (001) heterointerface that supports the strong Rashba-spin-valley states and highlighting the potential of the Bi$_{2}$O$_{2}$Se ML for future spintronics and valleytronics-based devices.  
\end{abstract}

\pacs{Valid PACS appear here}
\keywords{Suggested keywords}
\maketitle

\section{INTRODUCTION}

The correlation between the electron spin and orbital degrees of freedom is a fundamental concept in a range of emerging areas, such as spintronics and valleytronics \cite{Manchon, Schaibley, Ahn}. The examination of the symmetry of crystalline solids is critical in comprehending the physical properties of these new fields. For an instant, in a system that possesses both inversion and time-reversal symmetry, the doubly degenerate electronic bands are preserved throughout the first Brillouin zone (FBZ) even in the presence of the spin-orbit coupling (SOC), known as Kramer's spin degeneracy. When the inversion symmetry is broken, the SOC induces the momentum-dependent spin–orbit field (SOF) and lifts the band degeneracy through the so-called Rashba \cite{Rashba} and Dresselhaus \cite{Dress} effects. During the past decade, the Rashba-Dresselhauss effect has been the subject of intense research due to its potential applications in the emergent field of spintronics \cite{DiSante, Plekhanov}. In particular, the Rashba effect can manipulate the polarized spins by the external electric field \cite{Kuhlen} and thus holds promise for realizing spin field effect transistors (FETs) in experiments \cite{Datta}. 

In addition, an intriguing phenomenon arises when the time-reversal symmetry of crystalline solids is broken. In such cases, the absence of inversion symmetry, combined with the SOC, gives rise to the spin-valley contrasting effect \cite{Schaibley, Xiao2012}. Typically, the term valley refers to the energy extrema points in momentum space that represents a binary index for low-energy carriers, which is robust to the scattering by smooth deformations and long-wavelength phonons \cite{Xiao2012, Lu2013}. The valley degree of freedom is potentially utilized to store and carry information, leading to conceptual electronic applications known as valleytronics\cite{Gunawan, Rycerz2007}. Moreover, utilization of the spin-valley contrasting effect gives rise to topological
valley transport properties such as photo-induced charge Hall effect, valley Hall effect, and spin Hall effect under zero magnetic
field \cite{Xiao2012, Mak, Onga2017}. Several promising systems for hosting the spin-valley contrasting effect have been reported in two-dimensional (2D) systems, including the family of transition metal dichalcogenides (TMDCs) $MX_{2}$ ($M$=W, Mo; $X$=S, Se, Te) monolayer (ML) \cite{Xiao2012, Mak, Onga2017} and a class of septuple-atomic-layer MXenes $MA_{2}Z_{4}$ ($M$=W, Mo; $A$= Si, Ge; $Z$= N, P, As Te) ML \cite{Sheoran, Yang2021, Zhou2021}. Here, $D_{3h}$ symmetry of the TMDCs ML enforces the occurrence of the large spin splitting located at the $K$ valley in the FBZ, exhibiting valley-contrasting spin polarization perpendicular to the 2D plane. This effect has been evidenced by its unique optical and transport characteristics, including valley-dependent circular dichroic photoluminescence \cite{Zeng2012, Cao2012} and nonreciprocal charge transport \cite{Ryohei}.
 
In some systems with multiple well-separated valleys, the Rashba and spin-valley effects can both be observed. In such cases, coupling between the Rashba and spin-valley states can occur, resulting in Rashba-flavored spin-valley states. These states enable non-dissipative transport of both spin and valley degrees of freedom by utilizing the position of the Rashba and spin-valley states near the Fermi level. Typically, Rashba states produce a pair of spin-polarized bands in $k$-space with opposite chiral in-plane spin textures, which allows electrons to back-scatter and greatly reduces the efficiency of spin transport \cite{Dyakonov, Araki}. In contrast, back-scattering is suppressed in spin-valley states due to their unidirectional out-of-plane spin textures \cite{Liu2015, Yang2015}. The Rashba-spin-valley states have been observed in various 2D ML structures with in-plane mirror symmetry breaking, such as Janus TMDCs $MXY$ ($M$=W, Mo; $X, Y$=S, Se, Te) ML \cite{Yu2021, Absor2018, Yao2018}, Janus MXenes $MAA'Z_{2}Z'_{2}$ ($M$=W, Mo; $A, A'$= Si, Ge; $Z, Z'$= N, P, As Te) ML \cite{Rezavand, Hussain}, Janus Titanium dihalide Ti$XY$ ($X, Y$ = Cl, Br, I) ML \cite{Wang2019} and buckled hexagonal h-$M$N ($M$=V, Nb, and Ta) ML \cite{Ahammed}. However, the close energy proximity between the Rashba and spin-valley states in these systems may induce interference between in-plane and out-of-plane spin textures, resulting in low efficiency of spin transport. Recently, it has been reported that 2D layered centrosymmetric black arsenic also hosts Rashba-spin valley states that can be controlled by electrostatic gating \cite{Sheng2021}. However, the light atomic weight of arsenic leads to a small spin splitting of the Rashba-spin valley states.

This study utilizes first-principles density-functional theory (DFT) calculations to show that the Rashba-spin-valley states emerge in 2D bismuth oxyselenide, Bi$_{2}$O$_{2}$Se. Recently, 2D ultrathin layered Bi$_{2}$O$_{2}$Se has been successfully synthesized, exhibiting the strong SOC interaction \cite{Meng, Gao} and superior transport properties such as ultra-high Hall electron mobility \cite{Wu, Wu_A, ChengChen, KZhang}, making the achievement of Rashba-spin-valley states in this material significant for the development of spintronics and valleytronics applications. We find that breaking the crystal inversion symmetry of 2D Bi$_{2}$O$_{2}$Se ML using an external electric field leads to the Rashba-spin-valley formation exhibiting the Rashba-types splitting around the $\Gamma$ valley and spin-valley coupling at $D$ valleys located near the middle of the $\Gamma-M$ line. Unlike previous studies, unidirectional in-plane spin textures around the $D$ valleys are observed, in addition to conventional Rashba-type spin textures around the $\Gamma$ valley. These spin textures are driven by the electric field-induced lowering of crystal symmetry from $D_{4h}$ to $C_{4v}$ point group, as demonstrated through the $\vec{k}\cdot\vec{p}$ model derived from symmetry analysis. More interestingly, manipulating the Rashba and spin-valley states through biaxial strain offers an efficient means of managing the spin textures and spin splitting, thereby hindering electron back-scattering in spin transport. We further propose a more realistic system, namely, Bi$_{2}$O$_{2}$Se ML/SrTiO$_{3}$ (001) heterointerface, where the strong Rashba-spin-valley states is achieved. Since molecular beam epitaxy (MBE) of atomically thin Bi$_{2}$O$_{2}$Se film down to the ML structure grown on the SrTiO$_{3}$ (001) substrate has been experimentally reported \cite{Liang}, observation of the Rashba effect on the Bi$_{2}$O$_{2}$Se ML/SrTiO$_{3}$ (001) is expected to be realized in the near future. Therefore, the application of Bi$_{2}$O$_{2}$Se ML for spintronics and valleytronics devices is plausible.
 
\section{Model and Computational Details}

To evaluate the structural, electronic, and spin-splitting-related properties of Bi$_{2}$O$_{2}$Se ML, we performed fully relativistic DFT calculations, where the SOC was taken into account self-consistently by using total momentum ($j$)-dependent pseudo potentials \citep{Theurich}. We used norm-conserving pseudo-potentials and optimized pseudo-atomic localized basis functions in our DFT calculations implemented in the OpenMX code \cite{Ozaki, Ozakikino, Ozakikinoa}. We apply the generalized gradient approximation of Perdew, Burke, and Ernzerhof (GGA-PBE) \cite{gga_pbe, Kohn} as an exchange-correlation functional. The linear combinations of multiple pseudo atomic orbitals (PAOs) generated using a confinement scheme were used as the basis functions \cite{Ozaki, Ozakikino, Ozakikinoa}. A set of the PAOs basis functions was specified as Bi8.0-$s$3$p$2$d$2$f$1, Sr7.0-$s$3$p$2$d$2$f$1, Ti7.0-$s$3$p$2$d$2, Se7.0-$s$3$p$2$d$2, O6.0-$s$2$p$2$d$2, and H6.0-$s$2$p$2, where 8.0, 7.0, 7.0, 7.0, 6.0, and 6.0 are the cutoff radii (in bohrs) of Bi, Sr, Ti, Se, O, and H atoms, respectively. Here, the integers after $s$, $p$, $d$, and $f$ indicate the radial multiplicity of each angular momentum component. The accuracy of the basis functions, as well as pseudo-potentials we used, were carefully bench-marked by the delta gauge method \cite{Lejaeghere}. We make a periodic slab model with a sufficiently large vacuum layer (25 \AA) to avoid interaction between adjacent layers. We used an $8\times8\times1$ $k$-point and real space grids corresponding to energy cutoffs larger than 350 Ry to obtain the converged results of the self-consistent field (SCF) loops. The energy convergence criterion of $10^{-9}$ eV was used. Phonon dispersion band is used to evaluate the dynamical stability of the Bi$_{2}$O$_{2}$Se ML obtained by using ALAMODE code \cite{Tadano} based on the force constants obtained from the OpenMX code calculations.

To provide a more realistic system, we built a heterostructure where SrTiO$_{3}$ [001] substrate is applied on Bi$_{2}$O$_{2}$Se ML. Here, SrTiO$_{3}$ [001] surface was modeled using eight layers slab. Two stables formation of Bi$_{2}$O$_{2}$Se ML/SrTiO$_{3}$ [001] interface is considered, i.e., (i) Se-SrO and (ii) Se-TiO$_{2}$ interface termination. The stability of Bi$_{2}$O$_{2}$Se ML/SrTiO$_{3}$ [001] interface was evaluated by calculating the cohesive energy ($E_{\texttt{coh}}$) through the following relation,
\begin{equation}
\label{1a}
E_{\texttt{coh}}=\frac{E_{\texttt{Bi}_{2}\texttt{O}_{2}\texttt{Se}} + E_{\texttt{SrTiO}_{3}} - E_{\texttt{Tot}}}{A},             
\end{equation}
where $E_{\texttt{Tot}}$, $E_{\texttt{Bi}_{2}\texttt{O}_{2}\texttt{Se}}$, and $E_{\texttt{SrTiO}_{3}}$ are the total energy of Bi$_{2}$O$_{2}$Se ML/SrTiO$_{3}$ [001], Bi$_{2}$O$_{2}$Se ML surface, and isolate SrTiO$_{3}$ [001] surface, respectively, while $A$ represents surface area. 

To study the Rashba-spin valley effect, we applied a uniform external electric field on Bi$_{2}$O$_{2}$Se ML along the $z$-direction modeled by a sawtooth waveform during the SCF calculation and geometry optimization. We calculated the spin textures in the momentum $k$-space by deducing the spin vector components ($S_{x}$, $S_{y}$, $S_{z}$) in the reciprocal lattice vector $\vec{k}$ from the spin density matrix. By using the spinor Bloch wave function, $\Psi^{\sigma}_{\mu}(\vec{r},\vec{k})$, obtained from the DFT calculations after the self-consistent field (SCF) is achieved, we calculate the spin density matrix, $P_{\sigma \sigma^{'}}(\vec{k},\mu)$, by using the following equation \cite{Kotaka, Yamaguchi},  
\begin{equation}
\begin{split}
\label{1}
P_{\sigma \sigma^{'}}(\vec{k},\mu) &=\int \Psi^{\sigma}_{\mu}(\vec{r},\vec{k})\Psi^{\sigma^{'}}_{\mu}(\vec{r},\vec{k}) d\vec{r}\\
                                   &= \sum_{n}\sum_{i,j}[c^{*}_{\sigma\mu i}c_{\sigma^{'}\mu j}S_{i,j}]e^{i\vec{R}_{n}\cdot\vec{k}},\\
\end{split}
\end{equation}
where $S_{ij}$ is the overlap integral of the $i$-th and $j$-th localized orbitals, $c_{\sigma\mu i(j)}$ is expansion coefficient, $\sigma$ ($\sigma^{'}$) is the spin index ($\uparrow$ or $\downarrow$), $\mu$ is the band index, and $\vec{R}_{n}$ is the $n$-th lattice vector. This method has been successfully applied in our recent studies on various 2D materials\cite{Absor1, Absor2, Absor3, Absor4, Absor5, Absor6}.

\section{Results and Discussion}

\begin{figure*}
	\centering		
	\includegraphics[width=1.0\textwidth]{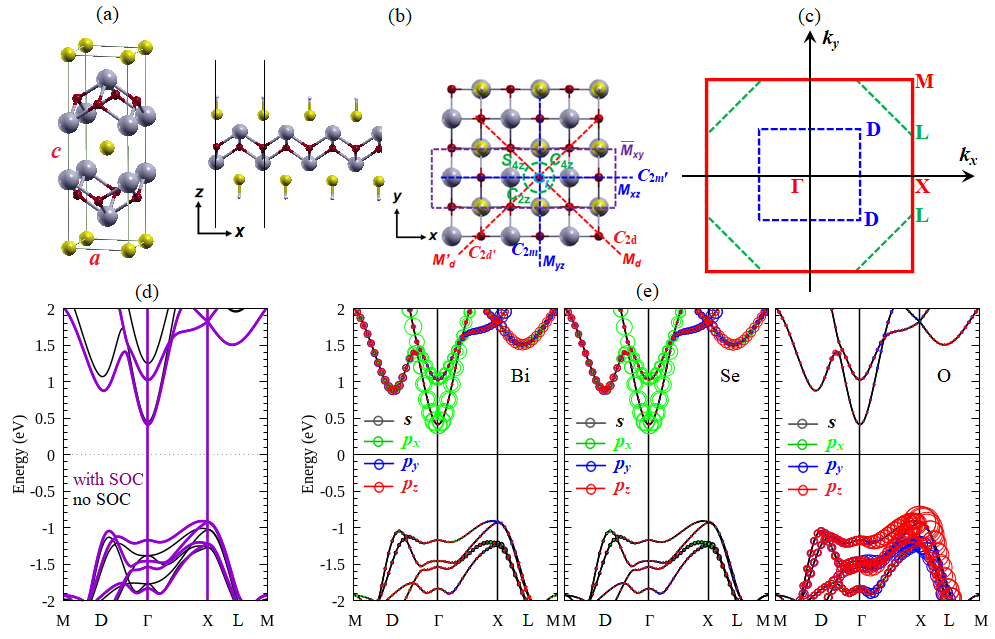}
	\caption{Atomic structures of Bi$_{2}$O$_{2}$Se for (a) bulk and (b) monolayer (ML) structures. For the case of the Bi$_{2}$O$_{2}$Se ML, hydrogen H-passivated configuration is applied for the upper and bottom surface Bi$_{2}$O$_{2}$Se ML to balance non-stoichiometry. The Grey, yellow, red, and white colors represent the Bi, Se, O, and H atoms, respectively. The symmetry operations of the Bi$_{2}$O$_{2}$Se ML,  including $E$, ${i}$, $C_{2z}$, $C_{2d,d'}$, $C_{2m,m'}$, $C_{4z\pm}$, $S_{4z\pm}$, $\overline{M}_{xy}$, $M_{xz,yz}$, and $M_{d,d'}$ are indicated. (c) Two-dimensional first Brillouin zone for both the Bi$_{2}$O$_{2}$Se ML characterized by time-reversal-symmetry points, such as $M$, $\Gamma$, and $X$ points, and non time-reversal-symmetry points, including $D$ and $L$ points, are highlighted. (d) electronic band structure of Bi$_{2}$O$_{2}$Se ML calculated without (black color) and with (magenta color) spin-orbit coupling (SOC). (e) Orbital-resolved projected bands for Bi, Se, and O atoms calculated with SOC are shown. The radii of the circles reflect the magnitudes of the spectral weight of the particular orbitals to the bands. }
	\label{figure:Figure1}
\end{figure*}

First, we examine the structural properties of Bi$_{2}$O$_{2}$Se ML. The geometry of the bulk and 2D ML structures are shown in Figs. 1(a)-(b), while the corresponding 2D FBZ is illustrated in Fig. 1(c). The layered Bi$_{2}$O$_{2}$Se crystallizes in a tetragonal structure with $I4/mmm$ space group [Fig. 1(a)], consisted of alternating stacking of positively charged [Bi$_{2}$O$_{2}$]$^{2n+}_{n}$ layers and negatively charged [Se$_{2}$]$^{2n-}_{n}$ layers with weak electrostatic interactions \cite{Wu}. The 2D ML structure of Bi$_{2}$O$_{2}$Se can be constructed by separating the crystal in the $c$ direction ($z$-direction) through the ionic bonds between [Bi$_{2}$O$_{2}$]$^{2+}$ and Se$^{2-}$ layers [Fig. 1(b)]. Here, the Se$^{2-}$ layers terminate both the top and bottom surface MLs, thus preserving the inversion symmetry. Similar to its bulk counterpart, Bi$_{2}$O$_{2}$Se ML belongs to the $D_{4h}$ point group symmetry \cite{Wu, LiangJ}, generated by the following symmetry operations: $E$, $i$, $C_{2z}$, $C_{2d,d'}$, $C_{2m,m'}$, $C_{4z\pm}$, $S_{4z\pm}$, $\overline{M}_{xy}$, $M_{xz,yz}$, and $M_{d,d'}$. Here, $E$ is the identity operator, $i$ is the inversion symmetry, $C_{2z}$ and $C_{4z\pm}$ represent the two-fold and four-fold rotations around the $z$ axis, respectively, $C_{2d, d'}$ and $C_{2m, m'}$ are the two-fold rotations around the dihedral and vertical mirror axis, respectively, $S_{4z\pm}$ is the four-fold rotations around the $z$ axis followed by reflection through the $xy$ plane, $\overline{M}_{xy}$ is the glide mirror symmetry through the $xy$ mirror plane followed by translation, $M_{xz,yz}$ is the vertical mirror symmetry through the $xz$ and $yz$ mirror plane, respectively, and $M_{d, d'}$ are the dihedral mirror symmetry containing the $z$ axis and crossing the angle between $k_{x}$ and $\pm k_{y}$ [Fig. 1(b)]. In our model, we applied hydrogen H-passivated configuration on Bi$_{2}$O$_{2}$Se ML for balancing the non-stoichiometry as used in the previous studies \cite{YZhang, Tang, WangN, Zhenqian}. We find that this structural configuration is dynamically stable as confirmed by phonon dispersion bands shown in Fig. S1 in the supplementary materials \cite{Supporting}. Nevertheless, the optimized in-plane lattice parameter of the ML (3.98 \AA) is a bit larger than that of the bulk (3.91 \AA), which agrees with prior findings \cite{YZhang, Tang, WangN, Zhenqian}. 

Fig. 1(d) depicts the electronic band structure of Bi$_{2}$O$_{2}$Se ML calculated with and without SOC. It demonstrates that the ML is a semiconductor having multiple valleys in its electronic band structures. Three valleys are observed at the conduction band, with the $\Gamma$ valley having the lowest energy, the $D$ valley located at the middle of the $\Gamma-M$ line, and the $L$ valley near the middle of the $X-M$ line [Fig. 1(d)]. On the contrary, the valence band is characterized by the $D$, $\Gamma$, and $X$ valleys, with the highest energy level located at the $X$ valleys. The $\Gamma$ valley in the conduction band mainly comes from the $p_{x}$ orbitals of the Bi and Se atoms, while the $D$ and $L$ valleys mostly originate from the mixing between $p_{x,y}$ and $p_{z}$ orbitals of the Bi and Se atoms [Fig. 1(e)]. On the contrary, $p_{z}$ orbital of the O atoms contributes dominantly to the $X$, $\Gamma$, and $D$ valley at the valence band. Since the conduction band minimum (CBM) and valence band maximum (CBM) is located at the $\Gamma$ and $X$ valleys, respectively, an indirect band gap is observed. The calculated value of the indirect band gap is 1.2 eV, which is in good agreement with previous theoretical reports \cite{YZhang, WangN}. We noted here that the indirect band gap of the 2D ultra-thin Bi$_{2}$O$_{2}$Se has been previously reported experimentally by using optical measurement \cite{Wu}. Although this indirect band gap is consistent-well with our calculations, the magnitude of the experimental band gap (1.95 eV) \cite{Wu} is larger than that of our result (1.2 eV), which is due to the use of the GGA-PBE in our DFT calculations. However, the topological band curvature including the position of the CBM and VBM obtained in our calculations is consistent-well with the ARPES experiment \cite{Wu, ChengChen}. When the SOC is taken into account, the indirect band gap of Bi$_{2}$O$_{2}$Se ML decreases to 1.1 eV. Since the inversion symmetry is preserved in Bi$_{2}$O$_{2}$Se ML, all the bands are spin degenerated [see red line in Fig. 1(d)]. Thus, there is no spin-splitting observed on Bi$_{2}$O$_{2}$Se ML.

In order to observe spin splitting in a Bi$_{2}$O$_{2}$Se ML, an external electric field ($E_{z}$) perpendicular to the ML surface is employed to break the crystal inversion symmetry. We utilized electric fields $E_{z}$ reaching a magnitude of 0.2 V/\AA, demonstrating thermodynamic stability as confirmed by the absence of imaginary frequency in the phonon dispersion; see Figs. S1 in the supplementary materials \cite{Supporting}. Figs. (2a)-(2c) illustrate evolution of the band structures of Bi$_{2}$O$_{2}$Se ML with different magnitudes of $E_{z}$, revealing that when $E_{z}$ is smaller than 0.18 V/\AA, the indirect band gap remains, but at larger $E_{z}$ of 0.18 V/\AA, it becomes metallic. This change from an indirect semiconductor to a metallic state, which is similar to that previously reported on InSe ML \cite{WangPeng2018} and bilayer phosphorene \cite{Feng2021}, is due to the strong coupling between the $p$-$p$ orbitals of Bi, Se, and O atoms in both the CBM and VBM [refer to Fig. 1(e)]. Additionally, the breaking of inversion symmetry together with SOC generates spin-splitting bands throughout the FBZ, except for high symmetry points at $\Gamma$, $X$, and $M$ valleys, which remain time-reversible [see Figs. 2(b)-(c)]. Since the spin splitting is more prominent in the bands near the Fermi level [Figs. 2(b)-(c)], hence this paper primarily focuses on the spin-split bands at both the CBM and VBM.

\begin{figure*}
	\centering		
	\includegraphics[width=1.0\textwidth]{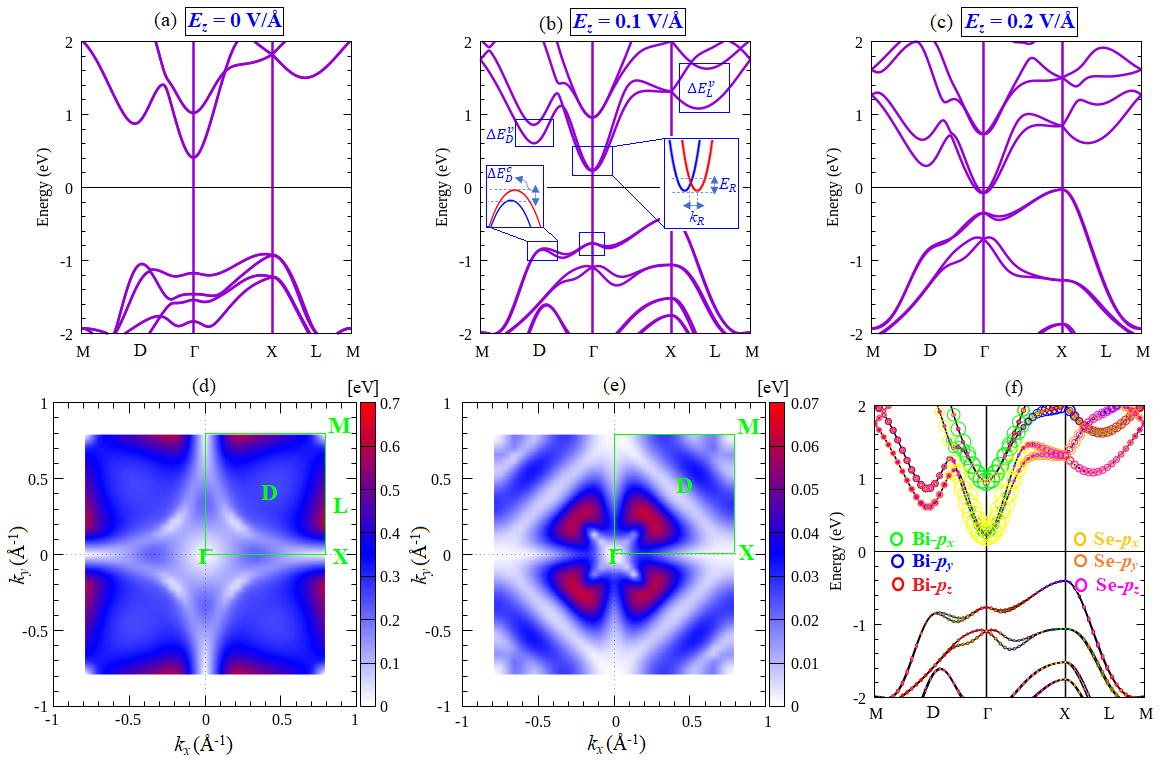}
	\caption{Evolution of the electronic band structure of Bi$_{2}$O$_{2}$Se ML calculated with the SOC under the out-of-plane electric field $E_{z}$: (a) 0 V/\AA, (b) 0.1 V/\AA, and (c) 0.2 V/\AA. Here, the spin-spin splittings at $D$, $\Gamma$, and $L$ valleys are highlighted. Spin-splitting energy of Bi$_{2}$O$_{2}$Se ML under $E_{z}$ of 0.1 V/\AA\ mapped on the first Brillouin zone calculated for (d) CBM and (e) VBM. The spin-splitting energy, $\Delta E(k)$, is calculated by using two bands at the CBM and VBM through the relation, $\Delta E(k)=\left\|E(k,\uparrow)-E(k,\downarrow)\right\|$, where $E(k,\uparrow)$ and $E(k,\downarrow)$ are the upper and lower bands of the spin-split states, respectively. (f) Orbital-resolved projected bands near the Fermi level for Bi and Se atoms are shown. The radii of the circles reflect the magnitudes of the spectral weight of the particular orbitals to the bands.}
	\label{figure:Figure2}
\end{figure*}

\begin{table*}
\caption{The spin-splitting parameters of the Rashba-spin-valley states calculated for the CBM [$\Delta E^{c}_{D, L}$ (eV), $\alpha^{c}_{R}$ (eV\AA)] and VBM [$\Delta E^{v}_{D}$ (eV), $\alpha^{v}_{R}$ (eV\AA)] of the Bi$_{2}$O$_{2}$Se ML under the influence of the $E_{z}$ of 0.1 V/\AA\ compared to those reported on the several 2D systems.} 
\begin{tabular}{c ccc ccc ccc ccc} 
\hline\hline 
  ML systems                  &&&  $\Delta E^{c,v}_{\texttt{valley}= D, L, K}$ (eV)  &&&           &&& $\alpha^{c,v}_{R}$ (eV\AA)   &&& Reference \\ 
\hline 
Bi$_{2}$O$_{2}$Se             &&&  0.32 ($D$ valley, CBM)    &&&                            &&&    1.33 (CBM)        &&&   This work \\
                              &&&  0.68 ($L$ valley, CBM)    &&&                            &&&                      &&&             \\
                              &&&  0.11 ($D$ valley, VBM)     &&&         &&&       0.52 (VBM)        &&&   \\
2D Janus TMDCs $MXY$               &&&                       &&&                              &&&               &&&    \\
MoSSe                          &&&  0.17 ($K$ valley)    &&&                                   &&&    0.07        &&&   Ref.\cite{Yu2021}\\
WSTe                          &&&  0.49 ($K$ valley)    &&&                                   &&&    0.48         &&&   Ref.\cite{Absor2018} \\
WSeTe                         &&&  0.45 ($K$ valley)    &&&                                   &&&    0.92         &&&   Ref.\cite{Yao2018} \\

2D Janus MXenes $MAA'Z_{2}Z'_{2}$  &&&                  &&&                                   &&&                &&&        \\
Mo(W)Si$_{2}$P$_{x}$As$_{y}$ ($x+y=4$)      &&&  0.14 - 0.51 ($K$ valley)          &&&                                    &&&   0.0 - 0.61        &&& Ref.\cite{Rezavand}\\
Mo(W)Ge$_{2}$P$_{2}$As$_{2}$                &&&  0.14 - 0.47 ($K$ valley)          &&&                                    &&&   0.5 - 0.52         &&& Ref.\cite{Hussain}\\

2D Janus TiXY ($X,Y$= Cl, Br, I)  &&&    0.04 - 0.06 ($K$ valley)              &&&                                &&&    0.07            &&&   Ref.\cite{Wang2019}     \\

Other 2D materials           &&&                       &&&                                      &&&                &&&    \\
BAs ML								       &&&     0.004 ($D$ valley)                  &&&                                      &&&                &&&   Ref.\cite{Sheng2021} \\
h-$M$N ($M$: V, Nb, Ta)     &&&  0.01 - 0.11 ($K$ valley)    &&&                                &&&    0.55 - 4.23            &&&  Ref.\cite{Ahammed} \\
\hline\hline 
\end{tabular}
\label{table:Table 3} 
\end{table*}

Figs. 2(d) and 2(e) depict the spin-splitting energy of a Bi$_{2}$O$_{2}$Se ML at an $E_{z}$ of 0.1 V/\AA\, mapped throughout the FBZ and computed at the CBM and VBM, respectively. Our findings reveal that the spin splitting shows a highly isotropic character around the center of the FBZ ($\Gamma$ valley), whereas it becomes strongly anisotropic near the edge of the FBZ ($X$ and $M$ valleys). More interestingly, it is evident that large spin splittings occur in both the CBM and VBM, which is particularly visible at the $D$ valley ($\Delta E^{c}_{D}=0.32$ eV; $\Delta E^{v}_{D}=0.11$ eV) and $L$ valley ($\Delta E^{c}_{L}=0.68$ eV), and there are also an apparent the significant Rashba-type spin splitting around the $\Gamma$ valley [refer to Figs. 2(b), 2(d), and 2(e)]. Due to time-reversibility at the $\Gamma$ valley, the spin splittings are dictated to have opposite signs at the $+D$ ($+L$) and $-D$ ($-L$) valleys, giving rise to an effective coupling between spin and valley pseudospin. In addition, due to the close in energy between the $\Gamma$ and $D$ valleys, the Rashba-spin-valley states are achieved. Our calculations confirmed that the strong coupling between the $p_{x,y}$ and $p_{z}$ orbitals of Bi and Se atoms is responsible for inducing the substantial spin splitting at the $D$ and $L$ valleys, as evidenced by the orbital-resolved projected spin-split bands shown in Fig. 2(f). On the other hand, the Rashba spin splitting around the $\Gamma$ valley can be quantified by using the Rashba parameter obtained from the linear Rashba model through the relation, $\alpha_{R}=2E_{R}/k_{R}$, where $E_{R}$ and $k_{R}$ are the Rashba energy and momentum offset, respectively [see the insert of Fig. 2(b)]. Then, we summarize the calculated results of the spin-splitting parameters ($\Delta E^{c,v}_{D}$, $\Delta E^{c}_{L}$, $\alpha^{c,v}_{R}$) in Table I, and compare these results with a few selected 2D materials supported Rashba-spin-valley states. Notably, the spin splittings at the $D$ ($\Delta E^{c,v}_{D}$) and $L$ ($\Delta E^{c}_{L}$) valleys are comparable with that observed in various 2D Janus TMDCs $MXY$ ML \cite{Yu2021, Absor2018, Yao2018} and Janus MXenes $MAA'Z_{2}Z'_{2}$ ML \cite{Rezavand, Hussain}, while the Rashba parameter $\alpha^{c,v}_{R}$ around the $\Gamma$ valley is much larger than those 2D Janus systems. Additionally, all of the splitting parameters ($\Delta E^{c,v}_{D}$, $\Delta E^{c}_{L}$, and $\alpha^{c,v}_{R}$) of the Rashba-spin-valley states are significantly larger than those observed on 2D Janus Ti$XY$ ($X,Y$= Cl, Br, I) ML \cite{Wang2019}, 2D buckled h-$M$N ($M$: V, Nb, Ta) ML \cite{Ahammed}, and 2D BAs ML \cite{Sheng2021}.

\begin{figure*}
	\centering		
	\includegraphics[width=1.0\textwidth]{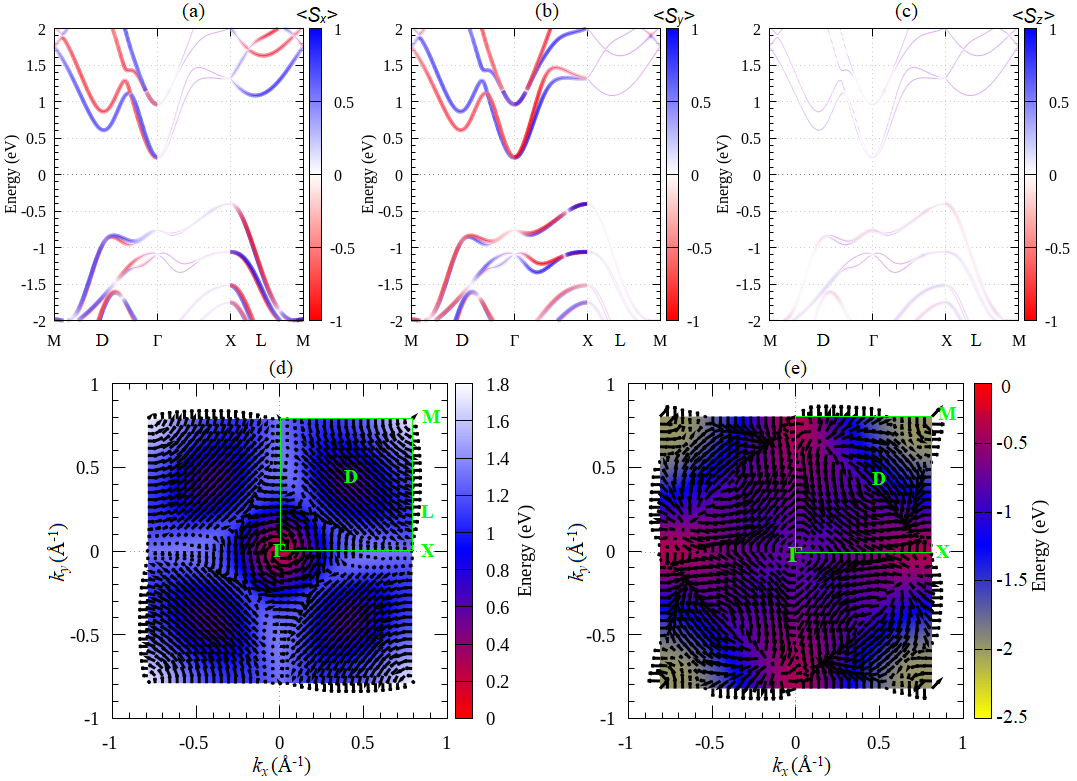}
	\caption{$k$-space spin component vector and spin texture of Bi$_{2}$O$_{2}$Se ML under the electric field $E_{z}$ of 0.1 V/\AA. The expectation value of (a) $S_{x}$, (b) $S_{y}$, and (c) $S_{z}$ spin components projected to the spin-split bands near the VBM is shown. (d)-(e) Energy-dependent of spin textures for the upper bands calculated near the CBM and VBM, respectively, are shown.}
	\label{figure:Figure3}
\end{figure*}

To gain a better understanding of the observed spin splitting of the Rashba-spin-valley states, we show in Figs. 3(a)-(c) expectation value of spin components ($S_{x}$, $S_{y}$, $S_{z}$) projected onto the spin-split bands near the Fermi level. Our findings show that the in-plane spin components ($S_{x}$, $S_{y}$) are the dominant contributors to the spin-split bands, while the out-of-plane spin components ($S_{z}$) make only a negligible contribution. By calculating the energy-dependent spin textures projected onto the FBZ, we observed a typical Rashba spin rotation for the spin textures around the $\Gamma$ valleys, whereas the spin textures become unidirectional pointing along the direction which is parallel to the $k_{x}-k_{y}$ plane around the $D$ valleys [Figs. 3(d)-3(e)]. In particular, the observed unidirectional in-plane spin textures persistently around the $D$ valley are strongly different from the spin textures of the spin-valley states observed on various 2D valleytronics materials where the fully-out-of plane spin textures are observed around the $K$ valley \cite{Xiao2012, Mak, Onga2017, Sheoran, Yang2021, Zhou2021, Yu2021, Absor2018, Yao2018, Rezavand, Hussain, Wang2019, Ahammed}. The observed unidirectional in-plane spin textures in the present system lead to the formation of the persistent spin textures \cite{Schliemann, Tao2018, Absor3, Absor4, Absor5, Absor6, Autieri2019}, which can prevent the electron from the back-scattering in spin transport and induces long-lived helical spin-wave mode through suppressing the Dyakonov spin-relaxation mechanism \cite{Dyakonov, Bernevig, Altmann}, which is promising for efficient spintronics.

\begin{table}[ht!]
\caption{Transformation rules for the wave vector $\vec{k}$ and spin vector $\vec{\sigma}$ under the $C_{4v}$ point-group symmetry.} 
\centering 
\begin{tabular}{cccc cccc cccc } 
\hline\hline 
  Symmetry operations       &&&& $(k_{x}, k_{y})$        &&&& $(\sigma_{x}, \sigma_{y}, \sigma_{z})$ \\
\hline 
$E$             &&&&  $(k_{x}, k_{y}, k_{z})$     &&&&  $(\sigma_{x}, \sigma_{y}, \sigma_{z})$  \\     
$C_{2z}$        &&&&  $(-k_{x}, -k_{y}, k_{z})$   &&&&  $(-\sigma_{x}, -\sigma_{y}, \sigma_{z})$ \\ 
$C_{4z+}$        &&&&  $(-k_{y}, k_{x}, k_{z})$   &&&&  $(-\sigma_{y}, \sigma_{x}, \sigma_{z})$ \\
$C_{4z-}$        &&&&  $(k_{y}, -k_{x}, k_{z})$   &&&&  $(\sigma_{y}, -\sigma_{x}, \sigma_{z})$ \\
$M_{d}$         &&&&  $(-k_{y}, -k_{x}, k_{z})$   &&&&  $(\sigma_{y}, \sigma_{x}, -\sigma_{z})$ \\
$M_{d'}$         &&&&  $(k_{y}, k_{x}, k_{z})$   &&&&  $(-\sigma_{y}, -\sigma_{x}, -\sigma_{z})$ \\
$M_{xz}$         &&&&  $(k_{x}, -k_{y}, k_{z})$   &&&&  $(-\sigma_{x}, \sigma_{y}, -\sigma_{z})$ \\
$M_{yz}$         &&&&  $(-k_{x}, k_{y}, k_{z})$   &&&&  $(\sigma_{x}, -\sigma_{y}, -\sigma_{z})$ \\
\hline\hline 
\end{tabular}
\label{table:Table 2} 
\end{table}

To clarify the origin of the observed spin splitting of Rashba-spin-valley states, we develop a two $\vec{k}\cdot\vec{p}$ band dispersion model derived from the symmetry analysis. This band dispersion can be determined by identifying all terms allowed by symmetry so that $H(\vec{k})=O^{\dagger}H(\vec{k})O$, where $O$ represents symmetry operations associated with the wave vector group ($G$) corresponding to the high-symmetry point and time-reversal symmetry. The invariant Hamiltonian should satisfy the condition given below \cite{Winkler},
\begin{equation}
\label{2a}
H_{G}(\vec{k})= D(O)H(O^{-1}\vec{k})D^{-1}(O),\ \forall O\in G, T  
\end{equation}
where $D(O)$ is the matrix representation of operation $O$ belonging to point group of the wave vector $G$. 

As mentioned previously that Bi$_{2}$O$_{2}$Se ML belongs to $D_{4h}$ PGS. When the electric field $E_{z}$ is applied, the symmetry of the crystal reduces to $C_{4v}$ point group, where the following symmetry operations, {$E$, $C_{2z}$, $C_{4z\pm}$, $M_{xz, yz}$, and $M_{d, d'}$}, remains. By using the transformation rules for the wave vector $\vec{k}$ and spin vector $\vec{\sigma}$ given in Table II, the symmetry allowed SOC Hamiltonian up to third order term is given by  
\begin{equation}
H_{\Gamma}(k)= H_{0}(k)+ \alpha (k_{x}\sigma_{y}-k_{y}\sigma_{x})+ \alpha^{'} k_{x}k_{y} (k_{y}\sigma_{y}-k_{x}\sigma_{x}) +\alpha^{"} (k_{x}^{3}\sigma_{y}-k_{y}^{3}\sigma_{x}),
\label{2}
\end{equation}
where $H_{0}(k)$ is the Hamiltonian of the free electrons with eigenvalues $E_{0}(k)=(\hbar^{2}k_{x}^{2}/2m_{x}^{*}) + (\hbar^{2}k_{y}^{2}/2m_{y}^{*})$, $m_{x}^{*}$ ($m_{y}^{*}$) is effective mass of electron evaluated from the band dispersion along the $k_{x}$ ($k_{y}$) directions, $k_{i}$ ($i = x, y$) are the components of the wave vector $k$ given regarding $\Gamma$ point taken as the origin, and $\sigma_{i}$ are the Pauli matrices. Here, $\alpha$ is the Rashba parameter which depends linearly on the electric field, $\alpha \approx \left|{E_{z}}\right|$. Moreover, the two last terms in Eq. (\ref{2}) which depend on $\alpha^{'}$ and $\alpha^{"}$ are the third-order terms in $k$, and are consistent with the derivation made by Vajna et al. \cite{Vajna} and Arras  et al. \cite{Arras}. The term $\alpha (k_{x}\sigma_{y}-k_{y}\sigma_{x})$ in Eq. (\ref{2}) has the usual form of the linear Rashba effect, which explains the isotropic in-plane Rashba spin rotation of the spin texture shown in Figs. 3(d)-(e). Moreover, near the $D$ valley, we find that $k_{x}\approx k_{y}=\widetilde{k}$, and we obtain that $H_{D}=\alpha \widetilde{k} (\sigma_{y}-\sigma_{x})$. This shows that the spin texture is oriented unidirectionally oriented in the in-plane direction parallel to the $k_{x}$-$k_{y}$ plane, which is also consistent-well with the observed spin textures around the $D$ valley presented in Figs. 3(d)-(e).

Solving the eigen-value problem involving the Hamiltonian of Eq. (\ref{2}), we obtain the spin-dependent eigenvalues [$E_{+}(k,\uparrow), E_{-}(k,\downarrow)$]. Accordingly, the splitting energy of the spin-split bands, $\Delta E(k)=\left\|E(k,\uparrow)-E(k,\downarrow)\right\|$, can be evaluated along the $\Gamma-X$ and $\Gamma-M$ lines as follows, 
\begin{equation}
\Delta E_{\Gamma-X}(k)= 2 (\alpha k_{x} + \alpha^{"} k_{x}^{3})
\label{3}
\end{equation}
and
\begin{equation}
\Delta E_{\Gamma-M}(k)= 2 \alpha k_{\|} + \widetilde{\alpha} k_{\|}^{3}),
\label{4}
\end{equation}
with $k_{\|}=\sqrt{k_{x}^{2}+k_{y}^{2}}$ and $\widetilde{\alpha}= \alpha^{'} + \alpha^{"}$. The parameters $\alpha$ and $\widetilde{\alpha}$ can be obtained by numerically fitting of Eqs. (\ref{3}) and (\ref{4}) to the spin splitting energy along the $\Gamma-X$ and $\Gamma-M$ lines obtained from our DFT calculations, respectively. We find that the calculated $\alpha$ and $\widetilde{\alpha}$ are 1.34 eV\AA (1.33 eV\AA) and 0.003 eV\AA$^{3}$ (0.01 eV\AA$^{3}$) for the spin-split bands at the CBM along the $\Gamma-X$ ($\Gamma-M$) line, respectively, while they are 0.524 eV\AA (0.517 eV\AA) and 0.01 eV\AA$^{3}$ (0.004 eV\AA$^{3}$) for the spin-split bands at the VBM along the $\Gamma-X$ ($\Gamma-M$) line, respectively. It is obvious that the obtained value of the cubic term parameters $\widetilde{\alpha}$ in both the $\Gamma-X$ and $\Gamma-M$ lines is too small compared with that of the linear-term parameter $\alpha$, indicating that the contribution of the higher-order correction is not essential. On the other hand, the calculated values of $\alpha$ which give an almost isotropic linear-Rashba parameter ($\alpha_{\Gamma-X}\approx \alpha_{\Gamma-M}$) obtained from the higher order correction is fairly agreement with that obtained from the linear Rashba model; see Table I.

\begin{figure*}
	\centering		
	\includegraphics[width=1.0\textwidth]{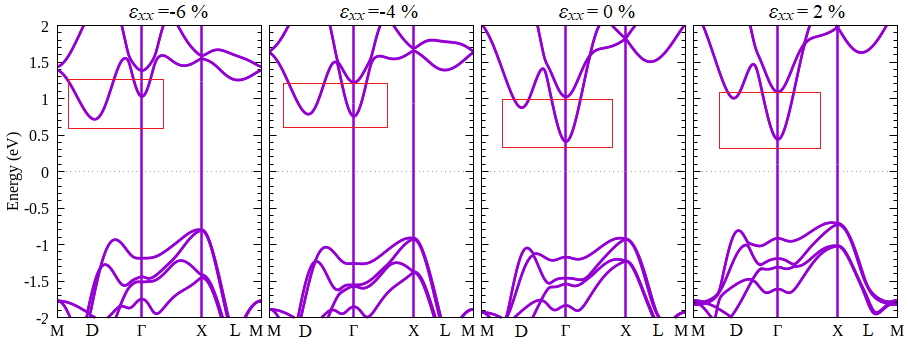}
	\caption{Strain-dependent of the electronic band structure calculated with the SOC is shown. The shift of the valley position in the CBM is highlighted.}
	\label{figure:Figure4}
\end{figure*}

Thus far, we have found that the Rashba-spin-valley state is observed in the Bi$_{2}$O$_{2}$Se ML, which is expected that this ML is suitable for spintronics. However, considering the different features of the spin textures around the $\Gamma$ and $D$ valleys [Figs. 3(d)-(e)], interference of the spin-polarized states may occur, which is not beneficial for spin transport. Although the appearance of the persistent spin textures observed around the $D$ valleys may hold non-dissipative spin transport \cite{Dyakonov, Bernevig, Altmann}, it may be disturbed by the Rashba spin texture around the $\Gamma$ valley owing to the back-scattering of electrons \cite{Dyakonov, Araki}. Therefore, suppressing the position of the Rashba states at the $\Gamma$ valley to be higher in energy than that of the spin-valley state at $D$ valley enables to prevent of the interference of the spin-polarized states. Since the multiple valleys observed near the band edges (CBM and VBM) are mostly characterized by the $p$-$p$ coupling orbitals [Fig. 1(e)], the valley positions in the electronic band structures can be effectively modified by the application of the biaxial strain. In fact, the valley-dependent strain has been previously reported on two-dimensional black phosphorene-type structures such as group IV monochalcogenide \cite{Anshory, Gomes}. Therefore, manipulating the Rashba-spin-valley states through biaxial strain offers an efficient route of managing the spin textures and spin splitting, thereby hindering electron back-scattering in spin transport. 

\begin{figure*}
	\centering		
	\includegraphics[width=1.0\textwidth]{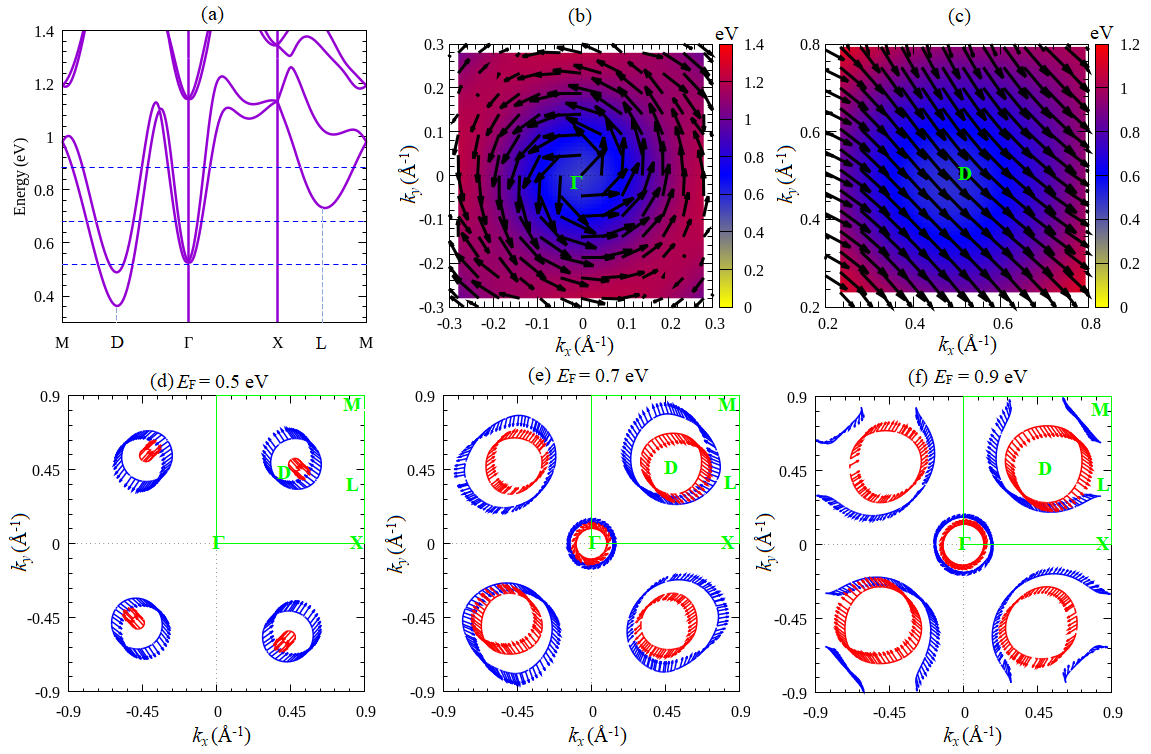}
	\caption{(a) The spin-split bands of the Bi$_{2}$O$_{2}$Se ML with $\epsilon_{xx}=-6$\% under the electric field $E_{z}$ of 0.2 V/\AA\ is shown. (b)-(c) Energy-dependent of the spin textures for the lower bands calculated around the $\Gamma$ and $D$ valleys near the CBM, respectively, is shown. (d) The spin textures calculated at different constant energy cuts of the Fermi line ($E_{\texttt{F}}$) for: (d) $E_{\texttt{F}}=0.5$ eV, (e) $E_{\texttt{F}}=0.7$ eV, and (f) $E_{\texttt{F}}=0.9$ eV, are presented.}
	\label{figure:Figure5}
\end{figure*}

We then introduce a wide range of biaxial strain (up to $\pm 10$\%), which is applied to the in-plane lattice constant of the Bi$_{2}$O$_{2}$Se ML. We define the degree of in-plane biaxial strain as $\epsilon_{xx}=\epsilon_{yy}=(a-a_{0})/a_{0}\times 100$\%, where $a_{0}$ is the unstrained in-plane lattice constant. Two different biaxial strains are studied, including the tensile strain, which
increases the in-plane lattice constant $a$, and the compressive strain, which decreases $a$. Our calculated results of the optimized structures revealed that these biaxial strains sustain the crystal symmetry and maintain the dynamical stability of the Bi$_{2}$O$_{2}$ML as evidenced by the phonon dispersion bands depicted in Fig. S1 of the supplementary materials \cite{Supporting}. The strain-dependent of electronic band structures of the Bi$_{2}$O$_{2}$Se ML under different strain conditions is shown in Fig. 4. Consistent with previous reports \cite{YZhang, WangN, Zhenqian}, it is revealed that the strained ML remains semiconductor at large strain up to $\epsilon_{xx}=\pm 10$\%. Under the tensile strain, an indirect band gap from $\Gamma$ valley (CBM) to $X$ valley (VBM) preserves similar to that of the equilibrium case. Conversely, the CBM starts to shift from the $\Gamma$ to $D$ valleys under the compressive strain larger than $4.8$\%. For the case of $\epsilon_{xx}=-6$\%, the $D$ valley is located much lower in energy than that of the $\Gamma$ valley at the CBM with a different energy of 0.42 eV. However, at the VBM, it is observed that the energy shift of the $D$ valley exceeds that of the $\Gamma$ valley by up to 0.23 eV. The decreasing (increasing) energy level of the $D$ valley with respect to the $\Gamma$ valley in the CBM (VBM) is expected to be useful for the spintronics since the spin texture of the spin-valley state around the $D$ valley could induce the spin-polarized states with minimal interference from the Rashba state around the $\Gamma$ valley.

\begin{figure*}
	\centering		
	\includegraphics[width=0.85\textwidth]{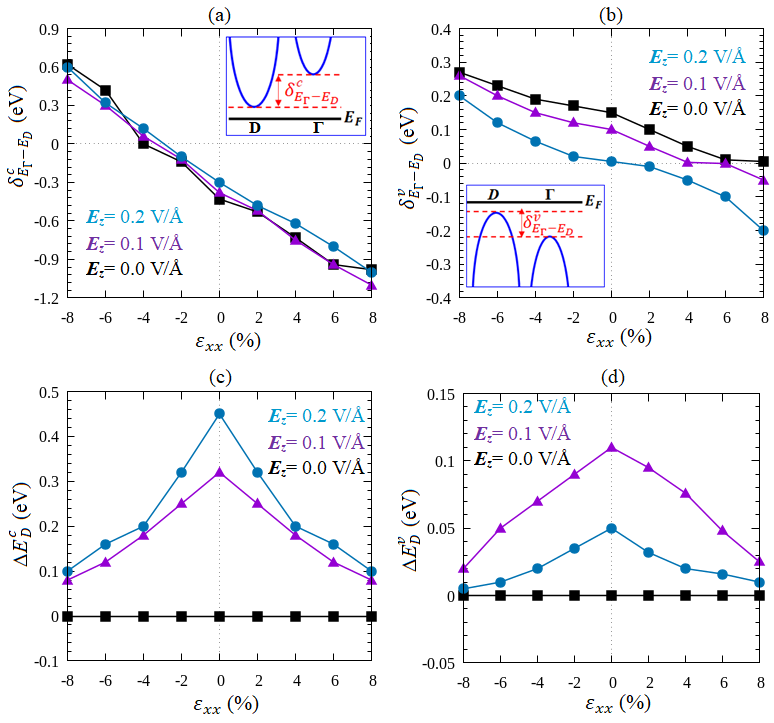}
	\caption{(a)-(b) Energy difference of the $\Gamma$ and $D$ valleys ($\delta^{c,v}_{E_{\Gamma}-E_{D}}=E_{\Gamma}-E_{D}$) as a function of biaxial strain for different strength of the external electric field $E_{z}$ calculated for the CBM and VBM, respectively. The insert shows the schematic view of the energy level of the $\Gamma$ and $D$ valleys with respect to the Fermi level to define $\delta^{c,v}_{E_{\Gamma}-E_{D}}$. (c)-(d) Strain-electric field-dependent of the spin-splitting energy at the $D$ valley calculated for the CBM and VBM is shown, respectively.}
	\label{figure:Figure6}
\end{figure*}

To clarify this, we further apply an external electric field $E_{z}$ of 0.2 V/\AA\ on the Bi$_{2}$O$_{2}$Se ML with $\epsilon_{xx}=-6$\% to observe the properties of the spin-polarized states. Concerning the large splitting bands at the CBM [Fig. 5(a)], we confirmed the presence of Rashba spin textures around the $\Gamma$ valley [Fig. 5(b)] and in-plane persistent spin textures around the $D$ valleys [Fig. 5(c)]. By comparing spin textures calculated at different constant energy cuts of the Fermi line ($E_{\texttt{F}}$) presented in Figs. 5(d)-(f), we clarified the possible interference of the spin-polarized states around the $\Gamma$ and $D$ valleys. At $E_{\texttt{F}}=0.5$ eV, all spin textures had a unidirectional in-plane orientation parallel to the $k_{x}-k_{y}$ plane, indicating that the spin-polarized states were achieved without any back-scattering. However, at larger $E_{\texttt{F}}$ [Figs. 5(e)-(f)], the mixing of spin textures around the $\Gamma$ and $D$ valleys resulted in the interference of spin-polarized states. Remarkably, non-dissipative spin transport can be achieved by tuning the position of the $E_{\texttt{F}}$ in the strained Bi$_{2}$O$_{2}$Se ML, which is important for operating spintronics devices.

Now, we discuss the strain-electric field-dependent of the $\Gamma$-$D$ valley positions as well as their spin-splitting energy. Figs. 6(a) and 6(b) show the energy difference of the $\Gamma$ and $D$ valleys with respect to the Fermi level ($\delta^{c,v}_{E_{\Gamma}-E_{D}}$) under the different strain and electric field [see the insert of Figs. 6(a)-(b) for the schematic view of $\delta^{c,v}_{E_{\Gamma}-E_{D}}$]. It is revealed that $\delta^{c,v}_{E_{\Gamma}-E_{D}}$ exhibit similar trends in both the CBM and VBM under different strain-electric field conditions. At the CBM, the positive value of $\delta^{c}_{E_{\Gamma}-E_{D}}$ is achieved under the compressive biaxial strain started from $\epsilon_{xx}=-4$\% and increases in magnitude under the increasing of the compressive biaxial strain and electric fields [Fig. (6a)]. Here, the CBM is mostly occupied by the $D$ valley states. Accordingly, the spin-polarized states are driven by the in-plane persistent spin texture suppressing the back-scattering of electrons and resulting in a highly efficient spin transport. On the other hand, the value of $\delta^{c}_{E_{\Gamma}-E_{D}}$ becomes negative when the tensile strain is applied, making the position of the CBM occupied by the $\Gamma$ valley state. Therefore, the spin-polarized states experience back-scattering due to the presence of the Rashba spin textures around the $\Gamma$ valley, and hence significantly reduces the efficiency of the spin transport. Since the similar trend of $\delta^{c}_{E_{\Gamma}-E_{D}}$ also holds for $\delta^{v}_{E_{\Gamma}-E_{D}}$ in the VBM [Fig. 6(b)], the similar character of the spin-polarized states is also expected. Considering the fact that the spin-splitting energy at the $D$ valley in both the CBM and VBM is tunable by the strain and electric field [Figs. 6(c)-6(d)], thus the present systems can be used as a promising platform for spintronic devices.

\begin{figure*}
	\centering		
	\includegraphics[width=1.0\textwidth]{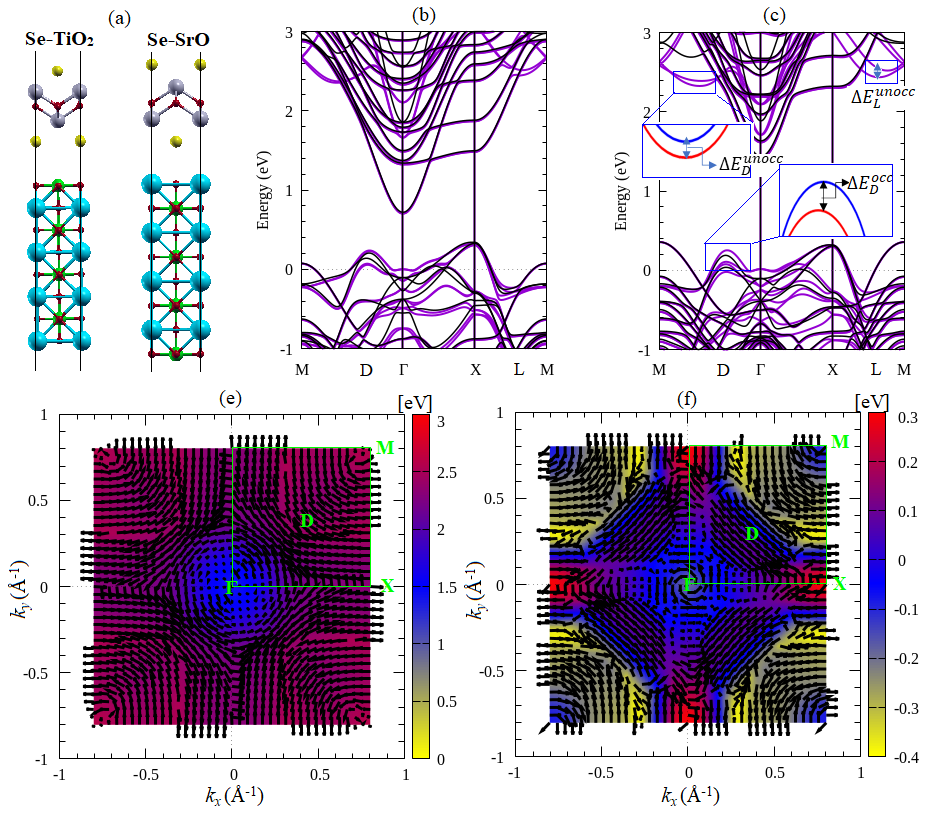}
	\caption{(a) Atomic structures of Bi$_{2}$O$_{2}$Se ML/SrTiO$_{3}$ [001] heterointerface with Se-SrO (left) and Se-TiO$_{2}$ (right) interface terminations are shown. The Grey, yellow, red, blue, and green colors represent the Bi, Se, O, Sr, and Ti atoms, respectively. The electronic band structures calculated with and without the SOC for Bi$_{2}$O$_{2}$Se ML/SrTiO$_{3}$ [001] heterointerface with (b)  Se-TiO$_{2}$ and (c) Se-SrO interface terminations, are given. The insert in Fig. 7(c) highlights the spin-splitting bands at the highest occupied states around the $D$ valley and the lowest unoccupied states around the $D$ and $L$ valleys. The calculated results of the energy-dependent of the spin textures projected to the FBZ for Bi$_{2}$O$_{2}$Se ML/SrTiO$_{3}$ [001] heterointerface with Se-SrO interface terminations calculated for (d) the lowest unoccupied and (e) highest occupied states are shown.}
	\label{figure:Figure7}
\end{figure*}

\begin{table*}
\caption{The spin-splitting energy at $D$ and $L$ valleys calculated at the highest occupied and lowest unoccupied states for Bi$_{2}$O$_{2}$Se ML/SrTiO$_{3}$ interface with different interface terminations is given. $\Delta E^{occ}_{D}$ (eV), $\Delta E^{unocc}_{D}$ (eV), and $\Delta E^{unocc}_{L}$ (eV) represent the spin splitting at $D$ valley for the highest occupied states, the lowest unoccupied states, and the spin splitting at $L$ valley for the lowest unoccupied state, respectively, while $\alpha^{occ}_{R,\Gamma}$ (eV\AA) and $\alpha^{unocc}_{R,\Gamma}$ (eV\AA) show the Rashba parameters around the $\Gamma$ point calculated for the highest occupied and lowest unoccupied states, respectively. } 
\begin{tabular}{ccc ccc ccc ccc ccc  ccc} 
\hline\hline 
  Termination systems    &&&  $\Delta E^{occ}_{D}$ (eV)  &&&  $\Delta E^{unocc}_{D}$ (eV)    &&& $\Delta E^{unocc}_{L}$ (eV)   &&&  $\alpha^{occ}_{R,\Gamma}$ (eV\AA)   &&&  $\alpha^{unocc}_{R,\Gamma}$ (eV\AA) \\ 
\hline 
         Se-SrO             &&&         0.11                &&&         0.09                    &&&    0.24      &&& 0.025    &&&  0.042   \\
         Se-TiO$_{2}$       &&&         0.0.08                &&&        0.04                    &&&    0.11     &&& 0.028    &&&  0.032    \\
\hline\hline 
\end{tabular}
\label{table:Table 3} 
\end{table*}

Finally, we explore the emergence of Rashba-spin-valley states in Bi$_{2}$O$_{2}$Se ML by considering a more realistic system. In this study, we propose an interfacial configuration consisting of a Bi$_{2}$O$_{2}$Se ML grown on a SrTiO$_{3}$ [001] substrate. We investigate two different interface terminations, namely Se-SrO and Se-TiO$_{2}$ terminations, as depicted in Figure 7(a). Both interface systems exhibit positive cohesive energies ($E^{\texttt{Se-SrO}}_{\texttt{coh}}=0.06$ eV/\AA$^{2}$ and $E^{\texttt{Se-TiO}_{2}}_{\texttt{coh}}=0.08$ eV/\AA$^{2}$), indicating a thermodynamically stable binding between the Bi$_{2}$O$_{2}$Se ML and SrTiO$_{3}$. Additional details, including the optimized interfacial lattice constant, interfacial distance, and interfacial charge transfer properties based on the average electrostatic potential, can be found in Table S1 and Fig. S2 in the supplementary materials \cite{Supporting}. Overall, our calculations confirm that the interfacial strength and stability of both interface systems are closely related to the bonding characteristics at the heterointerface.

Figs. 7(b)-7(c) show the electronic band structure of Bi$_{2}$O$_{2}$Se ML/SrTiO$_{3}$ interface with Se-SrO and Se-TiO$_{2}$ interface terminations calculated without (black) and with (red) including the SOC, respectively. Due to the defect contributor of the Bi$_{2}$O$_{2}$Se ML in the Bi$_{2}$O$_{2}$Se ML/SrTiO$_{3}$ interface, several occupied states near the Fermi level are observed, resulting in semi-metallic features of the electronic states. We find that the occupied states mainly originated from the Se-$p$ mixing with Bi-$6p$ states of the Bi$_{2}$O$_{2}$Se, while the unoccupied states are contributed mainly by the Ti-$d$ states with a small admixture of O-$2p$ and Sr-$d$ states of SrTiO$_{3}$, see Fig. S3 in the supplementary materials \cite{Supporting}. Importantly, we identify large spin splitting, which occurs around the $D$ and $L$ valleys as highlighted in Fig. 7(c). The calculated spin-splitting parameters ($\Delta E^{occ}_{D}$, $\Delta E^{unocc}_{D}$, $\Delta E^{unocc}_{L}$, $\alpha^{occ}_{R,\Gamma}$, $\alpha^{unocc}_{R,\Gamma}$) are listed in Table III for different interface terminations. Although these parameters are smaller than that of the Bi$_{2}$O$_{2}$Se ML, they are still comparable with that reported on various 2D materials listed in Table I. Moreover, these splittings exhibit highly persistent in-plane spin textures around the $D$ valley and the Rashba spin textures around the $\Gamma$ valley [see Figs. 7(d)-7(e)], indicating that the Rashba-spin-valley states is achieved similar to that observed on the Bi$_{2}$O$_{2}$Se ML. Due to a significant energy difference of approximately 0.22 eV between the $D$ valley and the $\Gamma$ valley in the highest occupied state, the dominance of in-plane persistent spin textures is expected to govern the spin-polarized states. This dominance serves as a protective mechanism against decoherence, facilitating non-dissipative spin transport. Given the experimental report by Liang et. al.\cite{Liang} on the successful molecular beam epitaxy (MBE) growth of an atomically thin Bi$_{2}$O$_{2}$Se film with an ML structure on the SrTiO${3}$ (001) substrate, it is expected that the observation of the Rashba-spin-valley states in the Bi$_{2}$O$_{2}$Se ML/SrTiO$_{3}$ (001) system will soon be achievable.

\section{Conclussion}

In summary, based on first-principles DFT calculations supported by symmetry analysis, we have systematically investigated the SOC-related properties of the strained Bi$_{2}$O$_{2}$Se ML. Breaking the crystal inversion symmetry of 2D Bi$_{2}$O$_{2}$Se ML with an external electric field causes the formation of Rashba-spin-valley states, characterized by Rashba-type spin splitting around the $\Gamma$ valley and spin-valley coupling at $D$ valleys along the $\Gamma-M$ line. In contrast to the Rashba-spin-valley states widely studied on previously reported 2D materials \cite{Xiao2012, Mak, Onga2017, Sheoran, Yang2021, Zhou2021, Yu2021, Absor2018, Yao2018, Rezavand, Hussain, Wang2019, Ahammed}, we observed persistent spin texture with in-plane orientation around the $D$ valleys, as well as conventional Rashba-type spin textures around the $\Gamma$ valley. The electric field-induced reduction of crystal symmetry from $D_{4h}$ to $C_{4v}$ point group is responsible for these spin textures, which is confirmed by a $\vec{k}\cdot\vec{p}$ model derived from symmetry analysis. Importantly, manipulating the Rashba and spin-valley states through biaxial strain provides an effective route to control the spin textures and spin splitting, which prevents electron back-scattering, and hence significantly enhances the efficiency of the spin transport.   

Since the Rashba-spin-valley states observed in the present study are solely enforced by the $C_{4v}$ point group symmetry, it is expected that these states can also be achieved on other 2D materials having a similar point group symmetry. Recently, another layered bismuth oxychalcogenide has been experimentally reported, including Bi$_{2}$O$_{2}$Te \cite{Ai2022} and Bi$_{2}$O$_{2}$S \cite{Yang2022}, which possess a similar crystal symmetry. Therefore, similar features of the Rashba-spin-valley states could be observed. 

Finally, the potential of utilizing the Bi$_{2}$O$_{2}$Se ML in spintronics and valleytronics-based devices is emphasized by considering a practical system, specifically the Bi$_{2}$O$_{2}$Se ML/SrTiO$_{3}$ (001) heterointerface. This system supports robust Rashba-spin-valley states, highlighting its relevance in realistic applications. Therefore, this prediction is expected to trigger further theoretical and experimental studies to clarify the emergence of the Rashba-spin-valley states in 2D-based bismuth oxychalcogenide systems, which would be useful for future spintronic and valleytronic applications.

\begin{acknowledgments}

This work was supported by collaborative research project 2023 supported by the Faculty of Mathematics and Natural Sciences, Universitas Gadjah Mada, Indonesia. The computation in this research was performed using the computer facilities at Universitas Gadjah Mada, Indonesia. 

\end{acknowledgments}

\bibliography{Reference1}


\end{document}